\documentclass{article}
\usepackage{amsmath, amssymb, amsfonts}
\title{ Time dependent embedding of a spherically symmetric Rindler-like spacetime } 
\author{Hristu Culetu, \\Ovidius University, Dept.of Physics, \\B-dul Mamaia 124, 900527 Constanta, Romania, \\e-mail : hculetu@yahoo.com}

\begin{document}
\numberwithin{equation}{section}
\pagenumbering{arabic}
\maketitle
\newcommand{\fv}{\boldsymbol{f}}
\newcommand{\tv}{\boldsymbol{t}}
\newcommand{\gv}{\boldsymbol{g}}
\newcommand{\OV}{\boldsymbol{O}}
\newcommand{\wv}{\boldsymbol{w}}
\newcommand{\WV}{\boldsymbol{W}}
\newcommand{\NV}{\boldsymbol{N}}
\newcommand{\hv}{\boldsymbol{h}}
\newcommand{\yv}{\boldsymbol{y}}
\newcommand{\RE}{\textrm{Re}}
\newcommand{\IM}{\textrm{Im}}
\newcommand{\rot}{\textrm{rot}}
\newcommand{\dv}{\boldsymbol{d}}
\newcommand{\grad}{\textrm{grad}}
\newcommand{\Tr}{\textrm{Tr}}
\newcommand{\ua}{\uparrow}
\newcommand{\da}{\downarrow}
\newcommand{\ct}{\textrm{const}}
\newcommand{\xv}{\boldsymbol{x}}
\newcommand{\mv}{\boldsymbol{m}}
\newcommand{\rv}{\boldsymbol{r}}
\newcommand{\kv}{\boldsymbol{k}}
\newcommand{\VE}{\boldsymbol{V}}
\newcommand{\sv}{\boldsymbol{s}}
\newcommand{\RV}{\boldsymbol{R}}
\newcommand{\pv}{\boldsymbol{p}}
\newcommand{\PV}{\boldsymbol{P}}
\newcommand{\EV}{\boldsymbol{E}}
\newcommand{\DV}{\boldsymbol{D}}
\newcommand{\BV}{\boldsymbol{B}}
\newcommand{\HV}{\boldsymbol{H}}
\newcommand{\MV}{\boldsymbol{M}}
\newcommand{\be}{\begin{equation}}
\newcommand{\ee}{\end{equation}}
\newcommand{\ba}{\begin{eqnarray}}
\newcommand{\ea}{\end{eqnarray}}
\newcommand{\bq}{\begin{eqnarray*}}
\newcommand{\eq}{\end{eqnarray*}}
\newcommand{\pa}{\partial}
\newcommand{\f}{\frac}
\newcommand{\FV}{\boldsymbol{F}}
\newcommand{\ve}{\boldsymbol{v}}
\newcommand{\AV}{\boldsymbol{A}}
\newcommand{\jv}{\boldsymbol{j}}
\newcommand{\LV}{\boldsymbol{L}}
\newcommand{\SV}{\boldsymbol{S}}
\newcommand{\av}{\boldsymbol{a}}
\newcommand{\qv}{\boldsymbol{q}}
\newcommand{\QV}{\boldsymbol{Q}}
\newcommand{\ev}{\boldsymbol{e}}
\newcommand{\uv}{\boldsymbol{u}}
\newcommand{\KV}{\boldsymbol{K}}
\newcommand{\ro}{\boldsymbol{\rho}}
\newcommand{\si}{\boldsymbol{\sigma}}
\newcommand{\thv}{\boldsymbol{\theta}}
\newcommand{\bv}{\boldsymbol{b}}
\newcommand{\JV}{\boldsymbol{J}}
\newcommand{\nv}{\boldsymbol{n}}
\newcommand{\lv}{\boldsymbol{l}}
\newcommand{\om}{\boldsymbol{\omega}}
\newcommand{\Om}{\boldsymbol{\Omega}}
\newcommand{\Piv}{\boldsymbol{\Pi}}
\newcommand{\UV}{\boldsymbol{U}}
\newcommand{\iv}{\boldsymbol{i}}
\newcommand{\nuv}{\boldsymbol{\nu}}
\newcommand{\muv}{\boldsymbol{\mu}}
\newcommand{\lm}{\boldsymbol{\lambda}}
\newcommand{\Lm}{\boldsymbol{\Lambda}}
\newcommand{\opsi}{\overline{\psi}}
\renewcommand{\tan}{\textrm{tg}}
\renewcommand{\cot}{\textrm{ctg}}
\renewcommand{\sinh}{\textrm{sh}}
\renewcommand{\cosh}{\textrm{ch}}
\renewcommand{\tanh}{\textrm{th}}
\renewcommand{\coth}{\textrm{cth}}

\begin{abstract}
An anisotropic cosmic fluid with radial heat flux which sources a time dependent Rindler-like geometry is investigated. Even though its energy density $\rho$ is positive, the radial and transversal pressures are negative and the strong energy condition is not satisfied. The congruence of ''static'' observers is not geodesic and the heat flux is oriented outward. We computed the Misner-Sharp energy associated with the Rindler-type metric embedded in a spatially flat FLRW universe and found that the Weyl energy is vanishing thanks to the conformally flat form of the spacetime. The null geodesic expansions are computed and one finds that only one of the two apparent horizons is located inside the event horizon. The properties of the Rindler-like geometry embedded in the conformally-flat de Sitter spacetime are investigated.

 \textbf{Keywords}: anisotropic cosmic fluid, MS energy, spherical Rindler metric, heat flux.
 
 \textbf{PACS}: 04.20.Cv, 04.70.Bw, 98.80.Jk\\
 \end{abstract}
 
\section{Introduction}
The apparently simple question whether the cosmological expansion takes place locally (in microphysics or at the Solar System level) is a very complicated one and is still an unsolved problem. The solution depends upon the model of the universe \cite{FJ, CG1, FMN, KS, SHMS}. The aim is to combine both classes of solutions (cosmological and local) and to find out exact solutions for the gravitational field of a particular system immersed in a cosmological background. However, a simple superposition of solutions is not in total agreement with the nonlinear feature of the theory \cite{CG2}. 

McVittie \cite{MV} was the 1st researcher taking into account the effect of the cosmic expansion on local gravitational systems (a central mass, for example). To avoid the accretion of the cosmic fluid into the central object, he imposed a constraint on its mass. However, the number and the existence of the apparent horizons (AHs) in McVittie spacetime depend on the cosmic time (for example, there are no AHs at early times \cite{FMN}). 

A more recent metric describing a spherical mass in a cosmological background is the Sultana - Dyer (SD) geometry \cite{SD}. Their line element is conformal to the Schwarzschild one and the stress tensor corresponds to two non-interacting perfect fluids (one timelike and the other null). Nevertheless, the cosmological fluid becomes tachyonic (negative energy density) at late times near the horizon. In addition, AHs do not exist at any time \cite{FJ}.

From a different point of view Mannheim \cite{PM} (see also \cite{DG}) obtained a fourth order Einstein's equations starting with a conformally-invariant Lagrangean - the Weyl tensor squared. These equations admit a static spherically-symmetric vacuum solution that contains, apart from the Schwarzschild term const.$/r$ a new term proportional to $r$ and, therefore, the metric is no longer asymptotically flat. According to Mannheim, this linearly rising potential term shows that a local matter distribution can actually have a global effect at infinity and so gravitational theories become global.

Saida et al. \cite{SHMS} studied black hole (BH) evaporation in an expanding universe (the so-called cosmological BH). In general the cosmological expansion will not influence the Hawking evaporation process if the cosmological horizon is much larger than the BH horizon. They investigated the no accretion case - the Einstein-Straus solution of Einstein's equation and the significantly accretion case - the SD solution (a conformally-transformed Schwarzschild metric), where a BH tends to increase its mass proportional to the cosmological scale factor. They found that the energy loss due to Hawking evaporation is negligible compared to the large mass accretion into the BH. 

Faraoni and Jacques \cite{FJ} examine various exact solutions representing strong field objects in a cosmological background. They showed that participation of a local object even strongly coupled (like a BH, for example) to the cosmological expansion seems to be the general rule. Therefore, Price's ''all or nothing'' behavior \cite{PR} is a peculiarity of the de Sitter spacetime used. In addition, the SD solution is, in their view, nonsingular at the horizon and can be interpreted as a BH embedded in a two-fluid universe. The same conclusion has been reached by Sun \cite{SHMS}, who wrote the correct explicit form of the scalar curvature. In contrast, Faraoni \cite{FJ} found a spacetime singularity at the event horizon of the same SD solution. The authors of \cite{FJ} further noticed that one is desirable to have cosmologic matter described by a single fluid and to drop the SD restriction on the special choice $a(t) \propto t^{2/3}$. In addition, Sun \cite{SHMS} calculated effectively the expansion of outgoing and ingoing future null vector fields normal to the spacelike 2-spheres foliating the two SD trapping horizons. 

Compared to the previous authors, we study in this paper how the cosmological expansion affects a local distribution of uniformly accelerated observers with spherical symmetry, described in Rindler-like coordinates. Following Sultana and Dyer, we introduce the cosmological expansion through a conformal factor, in terms of the scale $a(t)$. As we noticed before, Mannheim \cite{PM} already paid attention to the Rindler-type, linearly rising potential from the metric, related to a global effect of a local matter distribution. 

Our purpose is to analyze the structure of the stress tensor of the fluid which is the source of the metric curvature. Its components are functions of the scale factor $a(t)$ and the constant acceleration $g$. Moreover, we were interested to see whether the physical system contains a heat flux generated by a nonzero non-diagonal component of the energy-momentum tensor. On the other hand, an interesting feature of our spacetime relies on the existence of an event horizon (guaranteed by invariance of the causal structure to a conformal rescaling) and two AHs with simple expressions for their locations (as for the SD geometry). In contrast, a complicated sixth order equation for the AHs one encounters in the original McVittie metric or an eighth order equation for a generalized McVittie metric with a time varying mass \cite{GFSA}. A novel feature of our model is related to the expression of the Ricci part $E_{R}$ of the Misner-Sharp energy for an anisotropic fluid which, for $a(t) = 1$ acquires a previously obtained form of the Komar energy \cite{HC1}. Instead of studying a Schwarzschild BH embedded in a cosmological background wit $a(t) \propto t^{2/3}$ (in comoving time), as Sultana and Dyer did, or a BH immersed in a static de Sitter spacetime, we embed the Rindler-like spacetime in the same de Sitter space but written in conformally-flat coordinates, as we already did for the Schwarzschild BH \cite{HC2}. A local system immersed in a time dependent conformally-flat de Sitter universe has not been considered before, to the best of our knowledge.  

The paper is organized as follows. Sec.2 deals with the static form of the spherically-symmetric Rindler-type geometry, taking as the starting metric that one from \cite{HC1}. In Sec.3 we pay attention to the conformal time dependent version of the Rindler-like spacetime and keep track of the influence of the universe expansion on the local accelerating observers. The components of the energy-momentum tensor for such observers, the Misner-Sharp (MS) energy and the locations of the two AHs are calculated and analyzed. Sec.4 investigates the same Rindler observer embedded in a conformally-flat de Sitter universe and the consequences are evaluated. The last section sets the conclusions.

Throughout the paper we use geometrical units $c = G = 1$ and the positive signature $(-, +, +, +)$.   The Latin indices run from $0$ to $3$ and the coordinates order is ($t, r, \theta, \phi$).

\section{The spherically symmetric Rindler-type metric}
 We briefly review in this chapter the so-called ''Schwarzschild-Rindler metric'' \cite{HC1}, firstly obtained by Mannheim \cite{PM} and studied later by Grumiller \cite{DG}. Removing the central mass $M$, we reach the metric viewed, in our opinion, by a congruence of uniformly-accelerated observers located on an expanding sphere
   \begin{equation}
  ds^{2} = -(1- 2gr) dt^{2} + (1- 2gr)^{-1} dr^{2} + r^{2} d \Omega^{2}, 
 \label{2.1}
 \end{equation}
 where $g~>~0$ is the constant acceleration and $d \Omega^{2}$ stands for the metric on the unit 2-sphere. In \cite{HC1} the above (curved) metric has been applied in the interior of a relativistic star. To be a solution of the standard Einstein's equations 
   \begin{equation}
    \bar{G}_{ab} \equiv \bar{R}_{ab} - \frac{1}{2} \bar{g}_{ab} \bar{R}^{c}_{~c} = 8 \pi \bar{T}_{ab}   
 \label{2.2}
 \end{equation}
   one shows that a stress tensor is necessary on the RHS with nonzero components
   \begin{equation}
    \bar{T}_{0}^{0} = - \bar{\rho} = - \frac{g}{2 \pi r},~~~\bar{p}_{r} = \bar{T}_{1}^{1} = - \bar{\rho},~~~\bar{T}^{2}_{2} = \bar{T}^{3}_{3} = \bar{p}_{\bot} = \frac{1}{2} \bar{p}_{r},
 \label{2.3}
 \end{equation}
 where $\bar{\rho}$ is the energy density of the anisotropic fluid, $\bar{p_{r}}$ is the radial pressure and $\bar{p}_{\bot}$ represent tangential pressures. Hence, the spacetime (2.1) is sourced by an anisotropic fluid with negative pressure $\bar{p}_{r} = - \bar{\rho}$ and $\bar{p}_{\bot} = - \bar{\rho}/2$. It has an event horizon at $r = 1/2g$ and a true singularity at the origin $r = 0$, as can be seen from the scalar curvature and the Kretschmann scalar expressions
 \begin{equation}
  \bar{R}_{~a}^{a} = \frac{12g}{r}, ~~\bar{R}_{abcd} \bar{R}^{abcd} = \frac{32g^{2}}{r^{2}}
 \label{2.4}
 \end{equation} 
  We note that the Weyl tensor vanishes for the line element (2.1). In other words, it can be written in a conformally flat form (see \cite{HC5}). In addition, the trace of the energy-momentum tensor (2.3) is negative ($\bar{T}^{a}_{a} = -3g/2 \pi r$) and, as a consequence, the strong energy condition is not obeyed (as for the dark energy).
  
 \section{Conformal Rindler metric and universe expansion}
 In what follows we embed the geometry (2.1) in a spatially flat Friedmann-Lemaitre-Robertson-Walker (FLRW) universe written in a conformally flat form, by analogy with the Schwarzschild-de Sitter spacetime (a point source embedded in static de Sitter universe). Even though the Einstein equations are nonlinear we believe the above simple method may work in our situation.
   
  The ''combined'' metric appears as
   \begin{equation}
   ds^{2} = a^{2}(t) \left[-(1- 2gr) dt^{2} + (1- 2gr)^{-1} dr^{2} + r^{2} d \Omega^{2}\right], 
 \label{3.1}
 \end{equation}
 where $a(t)$ is the scale factor. We will dealing in this paper only with the domain $r < 1/2g$ (for $r > 1/2g$ a signature flip takes place and the radial coordinate becomes timelike). Let us note that, because of the time dependence of the metric (3.1), the event horizon $r = 1/2g$, obtained from the relation $g_{tt} = 0$, is not useful to be studied and we are more interested of the corresponding apparent horizon. We remark that the existence of the event horizon for the metric (3.1) is guaranteed because a conformal transformation does not affect the causal structure \cite{SHMS}. Its determinant does not tend to zero as $r \rightarrow 1/2g$. Hence, an object does not get crushed to zero volume as it approaches the singularity \cite{FJ}. Therefore, the singularity at $r = 1/2g$ is a weak one in the Tipler classification \cite{FT} and a particle could potentially cross a weak singularity. 
 
  We look now for the influence of the expansion ($\dot{a} \equiv da/dt >0$) of the universe on the local evolution of the system of uniformly accelerated observers that find themselves in the spacetime (3.1). Let us note that, far from the horizon ($r << 1/2g$) the line element becomes FLRW but when $a(t) = 1$, the spherical Rindler-type geometry is recovered. This is consistent with McVittie's prescription who combined Schwarzschild and FLRW metrics to produce a new metric that describes a point mass embedded in an expanding spatially-flat universe. As Carrera and Giulini \cite{CG1} have noticed, in a nonlinear theory as General Relativity there is no simple mathematical operation that produces a new solution out of two old ones (simple superposition of solutions is disallowed by nonlinearity). We are, however, tempted to use ''additivity'' of solutions that appears to be obvious due to the simultaneous physical presence of the Rindler acceleration and universe expansion. 
    
 Our next task is to impose (3.1) to be an exact solution of the Einstein equations
    \begin{equation}
    G_{ab} \equiv R_{ab} - \frac{1}{2} g_{ab} R^{c}_{~c} = 8 \pi T_{ab}.   
 \label{3.2}
 \end{equation}
To reach that purpose, $T^{a}_{~b}$ must have the nonzero components
 \begin{equation}
 \begin{split}
 -8 \pi T^{0}_{~0} = 8 \pi \rho =  \frac{3 \dot{a}^{2}}{a^{4}(1 - 2gr)} + \frac{4g}{a^{2}r},~~~8 \pi T^{0}_{~1} =  \frac{2g \dot{a}}{a^{3}(1 - 2gr)^{2}},\\ 
 8 \pi T^{1}_{~0} = - \frac{2g \dot{a}}{a^{3}},~~~  8 \pi T^{1}_{~1} = 8 \pi p_{r} = \frac{\dot{a}^{2} - 2a \ddot{a}}{a^{4} (1 - 2gr)} - \frac{4g}{a^{2}r},\\ 8 \pi T^{2}_{~2} = 8 \pi T^{3}_{~3} = 8 \pi p_{\bot} = \frac{\dot{a}^{2} - 2a \ddot{a}}{a^{4} (1 - 2gr)} - \frac{2g}{a^{2}r}.
 \end{split}
\label{3.3}
\end{equation}
The scalar curvature is given by
   \begin{equation}
 R^{a}_{~a} =  \frac{6 \ddot{a}}{a^{3} (1 - 2gr)} + \frac{12g}{a^{2}r}
 \label{3.4}
 \end{equation}
 Let us observe that $R^{a}_{~a}$ is divergent both at $r = 0$ and at the horizon $r = 1/2g$. The same is valid for the components of $T^{a}_{~b}$ from (3.3), excepting $T^{1}_{~0}$ and $T^{0}_{~1}$ which are finite at $r = 0$. When $a(t) = 1$ the expressions (2.3) are recovered. We have always $\rho > 0$ but the signs of $p_{r}$ and $p_{\bot}$ depend on the sign of $(\dot{a}^{2} - 2a \ddot{a}$). We remark the presence of an energy flux ($T^{1}_{~0}~ \neq 0$) which does not depend on the radial coordinate and becomes null when $g = 0$. In other words, the acceleration $g$ generates a time dependent flux of energy. 

Let us consider now a congruence of ''static'' observers with the velocity vector field
 \begin{equation}
   u^{a} = \left(\frac{1}{a(t) \sqrt{1 - 2gr}},~ 0,~ 0,~ 0\right),~~~u^{a} u_{a} = - 1 
 \label{3.5}
 \end{equation}
in the spacetime (3.1). One may check that the above congruence is not geodesic, the acceleration 4-vector being given by
 \begin{equation}
 a^{b} \equiv u^{a} \nabla_{a}u^{b} = \left(0,- \frac{g}{a^{2}},~0,~0\right),
 \label{3.6}
 \end{equation}
with $\sqrt{a^{b} a_{b}} = g/a \sqrt{1 - 2gr}$. The radial component of (3.6) does not depend on $r$ but the invariant acceleration diverges at the horizon. If one defines formally a surface gravity $\kappa$ as in the static geometries, one obtains
 \begin{equation}
   \kappa =  \sqrt{a^{b} a_{b}}~~ \sqrt{- g_{00}}|_{r = 1/2g} = g.
 \label{3.7}
 \end{equation}
However, we have to keep in mind that the geometry (3.1) is singular at $r = 1/2g$ and, therefore, the physical meaning of $\kappa$ is very doubtful. Moreover, as the radial acceleration $a^{r} < 0$, the force needed to keep the static observer at $r = const.$ is directed inward. In other words, the gravitational field is repulsive, as expected for a fluid with negative pressures.

As far as the scalar expansion of the congruence is concerned, we have
   \begin{equation}
  \Theta \equiv  \nabla_{a}u^{a} = \frac{3 \dot{a}}{a^{2} \sqrt{1 - 2gr}}, 
 \label{3.8}
 \end{equation}
 which is positive and diverges at the horizon. Its time evolution acquires the form
 \begin{equation}
\dot{\Theta} \equiv u^{a}\nabla_{a}\Theta = \frac{3 (-2\dot{a}^{2} + a \ddot{a})}{a^{4} (1 - 2gr)}.
\label{3.9}
\end{equation}
Taking into account the radial and transversal pressures are not equal and $T^{1}_{~0} \neq 0$, the source of the metric (3.1) is given by an anisotropic fluid with heat flux
  \begin{equation}
  T_{ab} = (p_{\bot} + \rho) u_{a} u_{b} + p_{\bot} g_{ab} + (p_{r} - p_{\bot}) s_{a}s_{b} +  u_{a} q_{b} + u_{b} q_{a},
 \label{3.10}
 \end{equation}
where $s^{a} = (0,~\sqrt{1 - 2gr}/a,~0,~0,)$ is a spacelike vector orthogonal to $u^{a}$, the energy density $\rho = T_{ab}u^{a}u^{b}$ and the heat flux is given by 
  \begin{equation}
  q^{a} = - T^{a}_{~b}u^{b} - \rho u^{a},~~~q_{a}u^{a} = 0.
 \label{3.11}
 \end{equation}
Eq. (3.11) yields
  \begin{equation}
  q^{a} = \left(0,~ \frac{g \dot{a}}{4 \pi a^{4}\sqrt{1 - 2gr}},~0,~0\right),
 \label{3.12}
 \end{equation}
with $q^{r} > 0$ and $q \equiv \sqrt{q^{a}q_{a}} = g \dot{a}/4 \pi a^{3} \left(1 - 2gr \right)$. It is worth noting that $q$ is finite at $r = 0$ but infinite at $r = 1/2g$. A comparison with Eq. (3.4) from \cite{HC2} shows that we have here $q^{r} > 0$ , i.e., the flux is oriented outward as if it were emanating from the central singularity. In addition, the constant acceleration $g$ from (3.12) plays the role of the Newtonian acceleration $m/r^{2}$ from Eq. (3.4) of \cite{HC2}. Since the metric (2.1) is not asymptotically flat, $q^{r}$ grows when $r$ increases and diverges at $r = 1/2g$.

In order to find energy $W$ that flows across a surface $\Sigma$ of constant $r$ for an arbitrary time interval, we have to integrate the heat flux (3.11), to obtain
  \begin{equation}
  W = \int (- T^{a}_{~b}u^{b} - \rho u^{a})n_{a}~\sqrt{- \gamma}~ d\theta~ d\phi~ dt,
 \label{3.13}
 \end{equation}
 where $n_{a} = (0,~a/\sqrt{1 - 2gr},~0,~0)$ is the normal to the hypersurface of constant $r$ and $\sqrt{- \gamma} = a^{3}\sqrt{1 - 2gr}~r^{2} sin\theta$, with $\gamma$ the determinant of the metric induced on the hypersurface $\Sigma$. Using now the expression of $T^{1}_{~0}$ from (3.3) and $u^{a}$ from (3.5), the last equation yields 
   \begin{equation}
  W = \frac{gr^{2}}{\sqrt{1 - 2gr}} \int^{t_{1}}_{t_{2}} \dot{a}(t)dt = \frac{gr^{2}}{\sqrt{1 - 2gr}} \Delta a(t)                                                        
  \label{3.14}
 \end{equation}
with $\Delta a(t) = a(t_{2}) - a(t_{1})$. Eq. (3.14) shows that the outgoing energy crossing $r = const.$ surface is positive (as required by the fact that $q^{r} > 0$), as if the fluid were attracted by the horizon $r = 1/2g$. For the outgoing flow rate we have 
   \begin{equation}
  \frac{dW}{dt} = \frac{gr^{2}\dot{a}}{\sqrt{1 - 2gr}}                                                        
  \label{3.15}
 \end{equation}
For $r << 1/2g$ (far from the horizon) one obtains $dW/dt \approx gr^{2}\dot{a} = m(r)\dot{a}(t)$, which is similar with the expression obtained by Faraoni and Jacques \cite{FJ} for the accretion rate (Hawking mass added per unit time), far from the horizon, in a generalized McVittie metric with arbitrary scale factor (their Eq. 84).

 We also note that $m(r) = gr^{2}$ plays the role of a mass as one already remarked in \cite{HC2}. For a small $\Delta t = t_{2} - t_{1}$, $\Delta a(t)$ is also small and $W$ as well. However, an observer located near $r = 1/2g$ could measure a large $W$, as if $\Delta t$ were large.

\section{Misner - Sharp energy and apparent horizons}
The Misner-Sharp (MS) quasilocal energy $E(t,r)$ \cite{CG1, NY, GGMP, CG2}, with its Weyl ($E_{W}$) and Ricci ($E_{R}$) parts, is useful to detect localized sources of gravity. In the case of spherical symmetry, $E$ is obtained from \cite{NY, CG2, HPFT} 
  \begin{equation}
  1 - \frac{2E(t,r)}{R} = g^{ab} \nabla_{a}R ~\nabla_{b}R,
 \label{4.1}
 \end{equation}
where $R = a(t)r$ is the areal radius and $\nabla_{a}R = \partial R/\partial x_{a}$. One finds, in the spacetime (3.1), that
  \begin{equation}
  E(t,r) = agr^{2} + \frac{\dot{a}^{2}r^{3}}{2a(1 - 2gr)}
 \label{4.2}
 \end{equation}
From the fact that the metric (2.1) (and therefore (3.1)) can be written in a conformally flat form, the Weyl tensor should vanish for the line element (3.1). Therefore, we get $E_{W} = 0$ \cite{CG2} because it is proportional to the Weyl scalar. In conclusion, $E_{R}$ must be given by (4.2). To show this, we remember that our fluid is not perfect and $E_{R}$ does not depend on $\rho$ only. One obtains, indeed, from
  \begin{equation}
  E_{R} = \frac{4 \pi}{3}R^{3}(\rho - p_{r} + p_{\bot}),
 \label{4.3}
 \end{equation}
and using the expressions from (3.3) for $\rho, p_{r}$ and $p_{\bot}$ , that
  \begin{equation}
  E_{R} = \frac{a^{3}r^{3}}{6} \left( \frac{3 \dot{a}^{2}}{a^{4}(1 - 2gr)} + \frac{6g}{a^{2}r}\right)
 \label{4.4}
 \end{equation}
which is exactly (4.2). One also observes that $E_{R}$ from (4.3) acquires the form $E_{R} = (4 \pi/3) R^{3} \rho$ when we are dealing with a perfect fluid ($p_{r} = p_{\bot}$). 

It is obvious that $E_{W} = 0$ (we have no a mass term in the metric). However, let us note that the 1st term on the RHS of (4.2) may be interpreted as a mass term if we write it as $a(t)m(r)$, with $m(r) = gr^{2}$ - a form already proposed in \cite{HC1} for the metric (2.1). 

The time dependence of the metric (3.1) makes the study of the AHs more appropriate compared to the event horizon. To locate them one must solve the equation \cite{HKS} 
  \begin{equation}
 g^{ab} \nabla_{a}R ~\nabla_{b}R = 0,
 \label{4.5}
 \end{equation}
whence one obtains
 \begin{equation}
  r_{\pm} = \frac{1}{2g \pm H},~~~~H \equiv \frac{\dot{a}}{a}.
 \label{4.6}
 \end{equation}
It is clear from (4.6) that the apparent horizon located at $r_{-}$ is not physical because $r_{-} = 1/(2g - H) > 1/2g$, namely outside our chosen domain (we must impose, of course, the condition $2g > H$ which is fulfilled with common accelerations $g$). The physical apparent horizon at $r_{+} = 1/(2g + H)$ is located below the singularity at $r = 1/2g$ and corresponds to the cosmological horizon at $r = 1/H$ when $g \rightarrow 0$ (or when $g << H$). The other singularity at $r = 0$ is naked, not being hidden by a horizon. 

We might also determine the location of the AHs by calculating the expansions of the outgoing and ingoing future-pointing radial null vector fields, $l^{a}$ and, respectively, $k^{a}$. The radial null geodesics for (3.1) coincide with those for (2.1) because the conformal factor does not change the causal structure. One obtains from (3.1)
 \begin{equation}
 \frac{dr}{dt} =\pm(1 - 2gr),
 \label{4.7}
 \end{equation}
where plus sign denotes the outgoing geodesics and minus sign - the ingoing ones. The components of $l^{a}$ and $k^{a}$ are given by
 \begin{equation}
 l^{a} = \frac{1}{a\sqrt{2(1 - 2gr)}}(1, 1 - 2gr, 0, 0),~~~k^{a} = \frac{1}{a\sqrt{2(1 - 2gr)}}(1, -1 + 2gr, 0, 0)
 \label{4.8}
 \end{equation}
and obey the relations $l^{a}l_{a} = 0,~k^{a}k_{a} = 0$. The temporal component $dt/d\lambda = 1/a\sqrt{2(1 - 2gr)}$, where $\lambda$ is the parameter along the null geodesic, has been set by means of the cross-normalisation $l^{a}k_{a} = -1$ \cite{HC4}. The expansions $\Theta_{l}$ and $\Theta_{k}$ associated with the vector fields $l^{a}$ and $k^{a}$ are given by
 \begin{equation}
 \Theta_{l} = \gamma^{ab}\nabla_{a}l_{b},~~~\Theta_{k} = \gamma^{ab}\nabla_{a}k_{b},
 \label{4.9}
 \end{equation}
where the projective tensor $\gamma^{ab} = (0, 0, 1/r^{2}, 1/r^{2}sin^{2}\theta)$ is the metric on a two dimensional surface to which the null geodesic congruence is orthogonal. It was obtained from
 \begin{equation}
 \gamma_{ab} = g_{ab} + l_{a}k_{b} + l_{b}k_{a}.
 \label{4.10}
 \end{equation}
Eqs. (4.9) and (4.10) yield
 \begin{equation}
 \Theta_{l} = a\sqrt{2(1 - 2gr)}\left[\frac{\dot{a}}{a(1 - 2gr)} + \frac{1}{r}\right],~~~\Theta_{k} = \frac{1}{a\sqrt{2(1 - 2gr)}}\left[\frac{\dot{a}}{a(1 - 2gr)} - \frac{1}{r}\right],
 \label{4.11}
 \end{equation}
where $\dot{a} = da/dt$. The AHs are located where $\Theta_{l}$ and, respectively, $\Theta_{k}$ vanish. Hence,
 \begin{equation}
 r^{-}_{AH} = \frac{1}{2g - H};~~~r^{+}_{AH} = \frac{1}{2g + H}.
 \label{4.12}
 \end{equation}
$r^{-}_{AH}$ is a past horizon since we have on it $\Theta_{k} = 0$ and $\Theta_{l} > 0$. In contrast, $r^{+}_{AH}$ is a future horizon since $\Theta_{k} < 0$ and $\Theta_{l} = 0$. We notice that the above results for the positions of the AHs are in accordance with (4.6). As already observed, the physical horizon is $r^{+}_{AH}$ because, as Nielsen \cite{AN} has proved, the AH (a marginally trapped surface) always lies inside the event horizon ($r = 1/2g$ in our case) and so cannot influence the outside region. In addition, we remind that for $r > 1/2g$ a signature flip will appear in (2.1) and the metric becomes time dependent.  

Let us notice the dependence of the AHs position on two independent parameters, $g$ and $H$. Even though the naked singularity is considered unphysical, we have in our situation two singularities (an inner one at $r = 0$ and an outer one at $r = 1/2g$). Perhaps there is some kind of compensation in the space between them and the spacetime is physical there.

\section{Time dependent Rindler-de Sitter geometry}
Our next task is to consider a particular choice of the scale factor $a(t)$. As in the previous paper \cite{HC2} (see also \cite{PD}) we choose the conformally flat de Sitter geometry as the space where the metric (2.1) is embedded, for to preserve the causal structure. Therefore, we take
  \begin{equation}
  ds^{2} = \frac{1}{H^{2}\eta^{2}} (-d \eta^{2} +  dr^{2} + r^{2} d\Omega^{2}) , 
 \label{5.1}
 \end{equation}
where $\eta$ is the conformal time and $H$ is the Hubble constant (noting that $H$ from Sec.4 is time dependent but in this section it is a true constant). $\eta$ is related to the cosmic time by $H \eta = - e^{- H \bar{t}}~ (\bar{t} > 0,~- 1/H < \eta < 0)$ \cite{HC2}). The ''combined '' spacetime appears as
  \begin{equation}
  ds^{2} = \frac{1}{H^{2}\eta^{2}} \left[- (1 - 2gr) d \eta^{2} +  \frac{dr^{2}}{1 - 2gr} + r^{2} d\Omega^{2}\right] , 
 \label{5.2}
 \end{equation}
where $a(\eta) = 1/H|\eta|$. Using this special value of the scale factor, we may write down the expression of the energy density and pressures of the anisotropic fluid
  \begin{equation}
  8 \pi \rho = - 8 \pi p_{r} = \frac{3H^{2}}{1 - 2gr} + \frac{4gH^{2}\eta^{2}}{r},~~~8 \pi p_{\bot} = - \frac{3H^{2}}{1 - 2gr} - \frac{2gH^{2}\eta^{2}}{r}
 \label{5.3}
 \end{equation}
It is easy to observe that the fluid becomes isotropic ($p_{r} = p_{\bot} \equiv p$) when $g = 0$ and, in addition, $\rho = - p = 3H^{2}$, as it should be for a de Sitter universe. Moreover, the isotropy is also obtained when $\eta \rightarrow 0$ (or $\bar{t} \rightarrow \infty)$. Nevertheless, in the latter case $\rho = - p = 3H^{2}/(1 - 2gr)$. We observe from (5.3) that the energy density and pressures become infinite at $r = 0$ or at the event horizon $r = 1/2g$. It is worth noting the positivity of $\rho$ for any values of $-1/H < \eta < 0$ and $r < 1/2g$. In contrast, $p_{r}$ and $p_{\bot}$ are always negative. The results are here more realistic than those obtained by Saida et al.\cite{SHMS}, where $\rho, ~p_{r}$ and $p_{\bot}$ change their sign for certain values of $r$ or $\eta$. Therefore, the weak energy condition is satisfied in the present case.

The expansion, scalar curvature and acceleration of the congruence are 
  \begin{equation}
  \Theta = \frac{3H}{1 - 2gr},~~~R^{a}_{~a} = \frac{12H^{2}}{1 - 2gr} + \frac{12gH^{2}\eta^{2}}{r},~~~a^{b} = \left(0,- gH^{2}\eta^{2},~0,~0\right).
 \label{5.4}
 \end{equation}
The expansion $\Theta$ is time independent ($\dot{\Theta} = 0$) and the curvature acquires the pure de Sitter value $12H^{2}$ when $g = 0$. For the heat flux one obtains
  \begin{equation}
  q^{r} = \frac{gH^{3}\eta^{2}}{4 \pi \sqrt{1 - 2gr}},~~~\sqrt{q^{a}q_{a}} = \frac{gH^{2}|\eta|}{4 \pi (1 - 2gr)} 
 \label{5.5}
 \end{equation}
For instance, taking the cosmic time $\bar{t} << H^{-1}$ (or $H|\eta| \approx 1$) and with $g \cong 10^{3} cm/s^{2}$, we have
  \begin{equation}
  q^{r} \approx \frac{c^{2}}{G}\frac{gH}{4 \pi} = 10^{12} erg/cm^{2}s
 \label{5.6}
 \end{equation}
We have used here $H \approx 10^{-18} s$ and $r << 1/2g$. The expression (5.6) resembles Eq. (4.10) from \cite{HC2} where instead of $g$ we have $m/r^{2}$, $m$ being the Schwarzschild mass. The constant flux (5.6) is outgoing ($q^{r} > 0$) and equals its value at the origin $r = 0$. 

\section{Conclusions}
We studied in this paper the (curved) Rindler-like space embedded in a spatially flat FLRW universe. In order to be an exact solution of Einstein's equations, the ''combined'' metric must be sourced by an anisotropic fluid with negative pressures and nonvanishing heat flux. The scalar invariants diverges at the origin of coordinates and at the horizon $r = 1/2g$, $g$ being the constant acceleration of a series of radially expanding observers (from the point of view of inertial observers). The off-diagonal components of the stress tensor become null when $g = 0$, that is the energy flux is generated by nonzero acceleration. Because of the anisotropy of the fluid, the Ricci part of the MS energy depends not only on $\rho$ but also on the pressures (the Weyl part vanishes thanks to the conformality of the metric). We finally studied the curved Rindler-like geometry embedded in a de Sitter space written in a conformally flat form. One obtains now a time independent scalar expansion of the congruence and for $g = 0$ the equation of state of the fluid acquires the form $\rho = - p = 3H^{2}$. So the pure de Sitter values are recovered.\\

\textbf{Acknowledgements}\\
The author gratefully thanks one of the anonymous referees for useful comments and suggestions which helped to improve the paper.


\begin{thebibliography} {23}

\bibitem{FJ}
V. Faraoni and A. Jacques, Phys. Rev. D76, 063510 (2007) (ArXiv: 0707.1350 [gr-qc]); V. Faraoni et al., Phys. Lett. B671, 7 (2009) (ArXiv: 0811.4667 [gr-qc]); V. Faraoni, Phys. Rev. D80: 044013 (2009) (ArXiv: 0907.4473 [gr-qc]).
\bibitem {CG1}
M. Carrera and D. Giulini, Phys. Rev. D81: 043521 (2010) (ArXiv: 0908.3101 [gr-qc]).
\bibitem{FMN}
V. Faraoni, A. Z. Moreno and R. Nandra, ArXiv: 1202.0719.
\bibitem{KS}
S. A. Klioner and M. H. Soffel, ArXiv: astro-ph/0411363.
\bibitem{SHMS}
H. Saida, T. Harada and H. Maeda, Class. Quant. Grav. 24: 4711 (2007) (ArXiv: 0705.4012 [gr-qc]); C.-Y. Sun, Comm. Theor. Ohys. 55, 597 (2011) (ArXiv: 1004.1760 [gr-qc).
\bibitem{CG2}
M. Carrera and D. Giulini, Rev. Mod. Phys. 82, 169 (2010) (ArXiv: 0810.2712 [gr-qc]).
\bibitem {MV}
G. C. McVittie, Mon. Not. R. Astr. Soc. 93, 325 (1933).
\bibitem{SD}
J. Sultana and C. C. Dyer, Gen. Relat. Grav. 37, 1349 (2005).
\bibitem{PM}
P Mannheim, Prog. Part. Nucl. Phys. 56, 340 (2006).
\bibitem{DG}
D. Grumiller, Phys. Rev. Lett. 105, 211303 (2010) (ArXiv : 1011.3625 [astro-ph]).
\bibitem{PR}
R. H. Price and J. D. Romano, arXiv: gr-qc/0508052.
\bibitem{GFSA}
D. C. Guariento, M. Fontanini, A. M. da Silva and E. Abdala, arXiv: 1207.1086 [gr-qc].
\bibitem{HC1}
H. Culetu, Int. J. Mod. Phys. Conf. Ser. 3, 455 (2011) (ArXiv: 1101.2980 [gr-qc]).
\bibitem{HC2}
H. Culetu, ArXiv: 1201.3769 [gr-qc].
\bibitem{HC5}
H. Culetu, arXiv: 1208.6450 [gr-qc].
\bibitem{FT}
F. J. Tipler, Phys. Lett. A64, 8 (1977).
\bibitem{NY}
A. Nielsen and D.-H. Yeom, Int. J. Mod. Phys. A24: 5261 (2009) (ArXiv: 0804.4435 [gr-qc]).
\bibitem{GGMP}
R. Giambo et al.,  Commun. Math. Phys. 235, 545 (2003) (ArXiv: gr-qc/0204030).
\bibitem{HPFT}
 L. Herrera et al., Int. J. Mod. Phys. D18: 129 (2009) (ArXiv: 0804.3584 [gr-qc]).
\bibitem{HKS}
S. Hellerman, N. Kaloper and L. Susskind, JHEP 0106: 003 (2001) (ArXiv: 0104180 [hep-th]). 
 \bibitem{HC4}
 H. Culetu, J. Korean Phys. Soc. 57, 419 (2010) (arXiv: 0905.3474 [hep-th]).
\bibitem{PD}
P. C. W. Davies, Quantum fields in curved spacetimes, Cambridge University Press (1982), pp. 130.
 \bibitem{AN}
 A. B. Nielsen, Gen. Rel. Grav. 41: 1539 (2009) (arXiv: 0809.3850 [hep-th]).




\end{thebibliography}
\end{document}